\documentclass[aps,prd,showpacs,twocolumn,amssymb]{revtex4}

\usepackage{dcolumn}
\usepackage{bm}
\usepackage{amsmath}
\usepackage{amsfonts}

\begin{document}
\title{Propagation of field disturbances in the Yang-Mills theory}
\author{Vitorio A. De Lorenci}
 \email{delorenci@unifei.edu.br}
\affiliation{Institute of Science, Federal University of Itajub\'a, 
37500-903 Itajub\'a, M. G., Brazil,}
\affiliation{PH Department, TH Unit, CERN, 1211 Geneva 23, Switzerland}
\author{Shi-Yuan Li}
 \email{lishy@sdu.edu.cn}
\affiliation{School of Physics, Shandong University, Jinan, 250100, P. R. China,}
\affiliation{PH Department, TH Unit, CERN, 1211 Geneva 23, Switzerland}

\date{\today}

\begin{abstract}
The propagation of field disturbances is examined in 
the context of the effective Yang-Mills Lagrangian, which is intended to be
applied to QCD systems. It is shown that birefringence phenomena can 
occur in such systems provided some restrictive conditions, as causality, 
are fulfilled. Possible applications to phenomenology are addressed. 
\end{abstract}

\maketitle


%
\section{Introduction}
Small disturbances on nonlinear fields propagate with velocity depending
on the polarization states. In general there will be two polarization
modes, leading to the  existence of two waves propagating with different velocities. This
phenomenon is known in the literature as birefringence. In the Maxwell theory 
(i.e., Abelian gauge field) it can be found when light propagates inside certain material media \cite{landau}. 
It can also appear in the context of nonlinear spin-one fields \cite{birula70,iacopini1979,delorenci00},
as it occurs in the quantum electrodynamics (QED). 
The effective Lagrangian for QED was derived long ago \cite{heisenberg36} for 
slowly varying but arbitrary strong electromagnetic fields. Its nonlinearities lead to effects 
like birefringence and  photon splitting \cite{birula70}. Some other investigations on this
issue can be found in \cite{rikken2000,adler2007,biswas2007,kruglov2007}.

For the case of non-Abelian gauge theories, the  
quantum fluctuations lead to a vaccum state which does not coincides with the vacuum
coming from the perturbation theories. The structure of the vacuum state 
was discussed for several models in 
\cite{savvidy1977,pagels78,nielsen1979,shuryak1980,ambjorn1980,ambjorn1980b}.
The one-loop effective action for Yang-Mills theory was
presented and discussed in, e.g., \cite{savvidy1977,pagels78}.
For the asymptotically free theory in the regime 
of large mean fields the effective action is controlled by perturbation theory. In 
this context the issue of event horizon formation in the 
physical vacuum associated with color confinement was considered in 
\cite{castorina07}. 

The mathematical formalism (see Section II for details)
to deal with the propagation of small disturbances in nonlinear
spin-one fields \cite{birula70,delorenci00} depends on the Lagrangian as a general
function of the field invariant.
Thus, it can be used to examine the wave propagation in systems governed by an Yang-Mills 
effective Lagrangian.
Particularly, it is worthwhile to analyze if effects like birefringence can occur in this 
context. Quantum Chromodynamics (QCD) could be taken as a specific application, since 
it presents strong nonlinear properties.

In this manuscript the one-parameter effective Lagrangian presented in \cite{pagels78}
is used as a `working model' when discussing Yang-Mills fields. 
Due to the possibility of two polarization modes presenting different velocities, 
as derived from our theoretical framework, it is shown that the birefringence phenomena can occur
provided that causal conditions are fulfilled.
In QCD case, we discuss 
how to observe the birefringence phenomena associated with the propagation of small
disturbances of the gluon field. Though the gluon is not directly 
observable due to confinement, a bulk of deconfined hot (and/or dense) quark-gluon 
matter is expected to exist in the ultra-relativistic heavy ion interactions, as well as 
in the early phase of the universe \cite{collins1975}. 
In those cases when the gluon propagates in the quark gluon matter,  it is argued that 
the birefringence of gluon field leads to local polarization of gluons. The polarization 
correlation is suggested to be measured in Gold-Gold collision on Relativistic 
Heavy Ion Collider (RHIC) at Brookhaven National Laboratory and Lead-Lead 
collision on Large Hadron Collider (LHC) at CERN. 

The paper is organized as follows.
In the Section II the light cone conditions associated with the propagation of 
small disturbances in one-parameter spin-one  theories are reviewed.  
In the Section III the effective Lagrangian for Yang-Mills
field \cite{pagels78} is presented and the procedure of taking volumetric spatial average is defined. 
Then it is discussed the non-trivial behavior of the phase velocity.
The conditions on the causal propagation are stated and some 
limiting cases from the effective Lagrangian are examined. 
Sections IV and V are  dedicated to the 
possible applications to phenomenology
and the effective geometry issue, respectively.
 Finally, some final remarks are presented in the 
conclusion section.

The present investigation is considered under the regime of the eikonal approximation, 
as addressed in \cite{delorenci00}. Latin indices run in the range $(1,2,3)$ 
and Greek indices run in the range $(0,1,2,3)$. The Minkowski spacetime 
is used, employing a Cartesian coordinate system. The background metric is denoted by
$\eta_{\mu\nu}$, which is defined by diag(+1,-1,-1,-1). Heaviside non-rationalized units 
are used and $c=1=\hbar$. The completely anti-symmetric tensor $\epsilon^{\alpha\beta\mu\nu}$ 
is defined such that $\epsilon^{0123} = 1$.

\section{Light cone conditions}
\subsection{Field equations for one-parameter spin-one theories}
The strength tensor field $F_{\mu\nu}^{(a)}$ and the gauge field $A_{\mu}^{(a)}$ are 
related by 
\begin{equation}
F_{\mu\nu}^{(a)} = \partial_{\mu}A_{\nu}^{(a)} -  \partial_{\nu}A_{\mu}^{(a)} 
+ C^{abc}A_{\mu}^{(b)}A_{\nu}^{(c)},
\label{Cabc}
\end{equation}
where $C^{abc}$ represent the structure constant for a compact Lie group $G$.
This tensor field can be conveniently defined in terms of the (non-Abelian)
electric $E_\mu^{(a)}$ and magnetic $H_\mu^{(a)}$  fields as
\begin{equation}
F_{\mu\nu}^{(a)} = V_{\mu}E_{\nu}^{(a)} -  V_{\nu}E_{\mu}^{(a)} - \epsilon_{\mu\nu}{}^{\alpha\beta}
V_{\alpha}H_{\beta}^{(a)},
\label{fmn}
\end{equation}
where $V_{\mu}$ represents the four-velocity of an observer at rest with respect to the 
laboratory. In Cartesian coordinates it is given by $V_{\mu}=\delta_{\mu}^{0}$. In order
to alleviate the notation, the `color' indices in the 
upper brackets will be omitted in what follows.

Let us assume the gauge invariant density of Lagrangian as a general function of the
Lorentz invariant $F\doteq F^{\mu\nu}F_{\mu\nu}$ as 
$L = L(F)$. From the minimal action principle we get the equation of motion
\begin{equation}
\label{9l}
\left(L_F F^{\mu\nu}\right){}_{,\nu} = 0.
\end{equation}
where a comma denotes partial derivatives with respect to the Cartesian coordinates. 
$L_F$ represents the derivative of $L$ with respect to the invariant $F$. $L_{FF}$ is
the second derivative. 
Using the relation $F_{,\nu} = 2F^{\alpha\beta}F_{\alpha\beta,\nu}$ 
in equation (\ref{9l}) we obtain:
\begin{eqnarray}
2
L_{FF}F^{\mu\nu}F^{\alpha\beta}
F_{\alpha\beta ,\nu} + L_F F^{\mu\nu}{}_{,\nu} = 0.
\label{19l}
\end{eqnarray}
The field strength $F_{\mu\nu}$ must satisfy the identity
\begin{equation}
\label{23l}
F_{\alpha\beta ,\lambda} + F_{\beta\lambda ,\alpha} + F_{\lambda\alpha ,\beta} = 0.
\end{equation} 
Let us now derive the expression for the light cone conditions for this class of
theories.

\subsection{The propagation of the field disturbances}
In this section we analyze the propagation of waves associated with
the discontinuities of the field \cite{Hadamard}. 
Let us consider a surface of discontinuity $\Sigma$ defined by
${\cal Z}(x^{\mu}) = 0$. Whenever $\Sigma$ is a global surface, it divides the spacetime 
in two distinct regions $U^-$ and $U^+$ (${\cal Z}<0$ and ${\cal Z}>0$, respectively). 
Given an arbitrary function of the coordinates, $f(x^\mu)$, 
we define its discontinuity on $\Sigma$ as
\begin{equation}
\label{24l}
\left[f(x^{\alpha})\right]_{\Sigma} \doteq \lim_{\{P^\pm\}\rightarrow P}
\left[f(P^+) - f(P^-)\right]
\end{equation}
where $P^+,\,P^-$ and $P$ belong to $U^+,\,U^-$ and $\Sigma$ respectively.
Applying the conditions \cite{Hadamard} for the tensor
field $F_{\mu\nu}$ and its derivative we set
\begin{subequations}
\begin{eqnarray}
\left[F_{\alpha\beta}\right]_{\Sigma} &=& 0 
\label{25al}
\\
\left[F_{\alpha\beta ,\lambda}\right]_{\Sigma} &=& f_{\alpha\beta}k_{\lambda}
\label{25bl}
\end{eqnarray}
\end{subequations}
where $f_{\alpha\beta}$ represents the discontinuities of field on
the surface $\Sigma$ and $k_\mu \doteq (\omega,\vec{k})$ represents the components 
of the wave 4-vector. The discontinuity of equations (\ref{19l}) and (\ref{23l}) 
yields, respectively,
\begin{eqnarray}
\label{34l}
&&f_{\beta\lambda} k^{\lambda} =
-\frac{2}{L_F}
L_{FF}F_{\beta}{}^{\mu}F^{\nu\rho}
f_{\nu\rho}k_{\mu},
\\
\label{26l}
&&f_{\alpha\beta} k_{\lambda} + f_{\beta\lambda}k_{\alpha} +
f_{\lambda\alpha} k_{\beta} = 0.  
\end{eqnarray}

For the case where $f_{\alpha\beta}$ is the wave propagation tensor 
given by equation (\ref{25bl}), for which equation (\ref{26l}) applies,
it follows that  
\begin{equation}
\label{fepsilon}
f_{\alpha\beta} = \sigma (\epsilon_\alpha k_\beta-\epsilon_\beta k_\alpha),
\end{equation}
where $\sigma$ is the strength of the wavelet and $\epsilon_\beta$
represents the polarization vector.  
Working with Eqs. (\ref{34l})-(\ref{fepsilon}) we obtain the eigenvalue
equation 
\begin{equation}
\label{Ze}
Z^\mu{}_\nu \epsilon^\nu = 0,
\end{equation}
where we defined
\begin{equation}
\label{Z}
Z^\mu{}_\nu \doteq \delta^\mu{}_\nu 
+\frac{4}{L_F k^2}
L_{FF}F^{\mu\alpha}F^{\nu\beta}
k_\alpha k_\beta,
\end{equation}
with $k^2\doteq k^\mu k_\mu$.  
The eigenvectors of $Z^\mu{}_\nu$ represent the dynamically allowed polarization 
modes $(e_+,e_-)$. 
The general solution for the eigenvalue equation  is formally given
by $det|Z^\mu{}_\beta|=0$, and results in the following light cone conditions 
\cite{boillat70,birula70,delorenci00}:
\begin{eqnarray}
k_+^2 &=& \gamma F^{\lambda\mu}F^{\nu}{}_{\lambda}k^+_{\mu}k^+_{\nu}, 
\qquad \gamma\doteq\frac{4L_{FF}}{L_F}
\label{+}
\\
k_-^2 &=& 0,
\label{-}
\end{eqnarray}
The $\pm$ signs are related with the two possible polarization
modes associated with the wave propagation \cite{birula70}. 
The existence of these two solutions shows that birefringence effects 
may appear, provided that $L_{FF}/L_F\neq 0$.
In the formalism of geometrical optics it is usually said that generally there will be two
rays inside the medium, the ordinary ray ($o$-ray) and the extraordinary ray ($e$-ray).
The former does not depend on the direction of wave propagation and its velocity
is equal to the light velocity in the classical vacuum of electrodynamics.
The latter presents an explicit dependence on the direction of propagation.
The light cone conditions for two-parameters Lagrangians can be obtained
in the same lines. For further details see Refs. \cite{birula70,delorenci00}. 

\section{Wave propagation in the Yang-Mils field}

\subsection{The effective Lagrangian}
The effective Lagrangian for QCD in terms of the parameter background field
$F$ can be presented \cite{pagels78} in the form 
\begin{eqnarray}
{\cal L}_{eff} = \frac{1}{4}\frac{F}{\bar{g}(t)^2}, \qquad t\doteq \log\frac{F}{\mu^4} 
\label{1}
\end{eqnarray}
where the effective coupling $\bar{g}(t)$ is implicitly given by
\begin{equation}
t = \int_{g}^{\bar{g}(t)}\!\!\! dg\; \frac{1}{\beta(g)},
\label{73}
\end{equation}
with $\beta(g)$ the Callan-Symanzik $\beta$-function and $g$ the gauge field coupling
constant appearing in the basic QCD Lagrangian.

In fact, there are many invariants of the Yang-Mills field
with the number dependent on the specific gauge group \cite{roskies1977}. 
The ansatz used to derive this effective Lagrangian takes in consideration only the algebraic 
invariant $F$ and imposes consistency with the trace anomaly for the energy-momentum tensor
\cite{collins1977}.

For the present proposes the system described by Eq. (\ref{1}) is assumed to  
satisfy the following requirements:
\begin{enumerate}
\item
the volumetric spatial average of the color field strength is independent of direction;
\item
it is equally probable that the products ${E^i E^j}$ and 
${H^i H^j}$ (with $i \neq j$), at any time, take positive or negative 
values;
\item
there is no net flow of energy as measured by a observer at rest with respect to the system. 
\end{enumerate}
The above mentioned volumetric spatial average of an arbitrary quantity $X$ 
for a given instant of time $t$ is defined as  
\begin{equation}
\label{Ref-1}
\overline{X} \stackrel{.}{=} \lim_{V\rightarrow V_o}
\frac{1}{V}\int X\,\sqrt{-g}\,d^3\!x^i,
\end{equation}
with $V=\int\sqrt{-g}\,d^3\!x^i$, and $V_o$ stands for the time dependent volume of the whole space.  
Similar average procedure has been already considered in the context of general relativity
solutions \cite{tolman1930,delorenci2002,kunze2008}.

In terms of the color fields, these requirements imply that:
\begin{eqnarray}
\overline{\rule{0pt}{2ex}E_i} = 0,\qquad
%
\overline{\rule{0pt}{2ex}H_i} &=& 0,\qquad
%
\overline{\rule{0pt}{2ex}E_i\, H_j - H_i\, E_j} = 0,
\label{meanEH}\\[1ex]
\overline{\rule{0pt}{2ex}E_i\,E_j} &=& -\, \frac{1}{3} E^2 \,\eta_{ij},
\label{meanE2}\\[1ex]
\overline{\rule{0pt}{2ex}H_i\, H_j} &=&  -\, \frac{1}{3} H^2 \,\eta_{ij},
\label{meanH2}
\end{eqnarray}
where we have defined $E^2\doteq -E^i E_i$ and $H^2\doteq -H^i H_i$.
The above average procedure consists in an idealization to deal with
systems like a quark-gluon plasma. Nevertheless, it has
adequate elements for our discussions.  
From the point of view of a statistical ensemble, we can assume that 
the field average over the whole bulk is vanishing comparing to the 
fluctuation, so it is isotropic as whole, while anisotropic for each local area. 

\subsection{Application to the effective Yang-Mills Lagrangian}
For the special Lagrangian presented in Eq. (\ref{1}), the factor $\gamma$ in 
Eq. (\ref{+}) is given by
\begin{equation}
\gamma = \frac{-4}{E^2(Z^2-1)}G(\bar{g}),
\end{equation}
where we have defined the quantities:
\begin{eqnarray}
Z^2 & \doteq & \frac{H^2}{E^2}
\label{86} 
\\
G(\bar{g}) & \doteq & \frac{\bar{g}\dot{\bar{g}} - 3\dot{\bar{g}}^2 
+ \bar{g}\ddot{\bar{g}}}{\bar{g}^2-2\bar{g}\dot{\bar{g}}},
\label{87}
\end{eqnarray}
with $\dot{\bar{g}} \doteq \partial \bar{g} / \partial t$.

The phase velocity for the wave perturbation can be obtained from the dispersion relation
as $v_e^2=\omega / |\vec{k}|$, where the index $e$ stands for the $e$-ray.
The $o$-ray propagates with the light velocity, as determined by Eq. (\ref{-}).
Now using the previous results, we obtain
\begin{equation}
v_e^2= 1 - \frac{8}{3}\frac{(Z^2+1)G(\bar{g})}{(Z^2-1)+4G(\bar{g})}.
\label{88}
\end{equation}
In order to guarantee causality, the physical solutions must 
satisfy the requirement  $0 \le v_e \le 1$, which implies that:
\begin{equation}
0 \le \frac{8}{3}\frac{(Z^2+1)G(\bar{g})}{(Z^2-1)+4G(\bar{g})} \le 1.
\label{89}
\end{equation}
From the analysis of the energy density for the effective action associated with
this problem, it can be inferred
that the case $E^2>H^2$ leads to a metastability of the vaccum. The
interpretation for this behavior is that if a region in the system develops a large
$E$ field, it will decay fastly to a configuration where $H^2>E^2$ \cite{pagels78}. 
Therefore it is adopted here that $H^2>E^2$ (which means $Z^2>1$ and $F>0$) and the 
condition stated by Eq. (\ref{89}) yields in 
\begin{equation}
0 \le G(\bar{g}) \le \frac{3(Z^2-1)}{4(2Z^2-1)}. 
\label{92}
\end{equation}

Now we are going to examine two cases in which the explicit form of the effective coupling
was presented in the literature. The first one is obtained when the regime of small 
coupling is taken in consideration, and the second one was previously proposed in \cite{gross1973}
as a suggestion for the case of large coupling constant.
%
%

For the case of small coupling the beta function can be expanded as
\cite{gross1973,poli1973,jone1974,casw1974}
\begin{equation}
\beta(g) = -\frac{1}{2}b_0 g^3 + b_1 g^5 + \cdots
\label{93}
\end{equation}
where $b_0$ and $b_1$ are constants. Now, taking the limit of large mean fields ($F\rightarrow\infty$)
we obtain from Eq. (\ref{73}) that
\begin{equation}
\frac{1}{\bar{g}(t)} = b_0 t - 2\frac{b_1}{b_0} \log t + \cdots
\label{94}
\end{equation}
Introducing these results in the effective Lagrangian we obtain \cite{pagels78,nielsen1978}
\begin{eqnarray}
{\cal L}_{eff} = \frac{1}{4}b_0 F \log\frac{F}{\mu^4}.
\label{95}
\end{eqnarray}
For this case the function $G(\bar{g})$ results to be
\begin{equation}
G(\bar{g}) = -\frac{1}{2} \frac{1}{t+1}.
\label{99}
\end{equation}
Since $t>1$ at the large mean field regime we conclude that $G(\bar{g})<0$, 
and there will be no propagation associated with the $e$-ray.
The $o$-ray travels with the unperturbed velocity $v_-=1$ and does not depend on the
direction of propagation. 

%
%
The non-perturbative expression for the beta function is still unknown. 
Nevertheless, a suggestion about the strong coupling form of the
beta function was presented long ago \cite{gross1973}, and is given
by
\begin{equation}
\beta(g) = -\frac{a}{2} g,
\label{100-1}
\end{equation}
with $a$ a positive constant. If we assume this form for the beta
function, we obtain from Eqs. (\ref{73}) and (\ref{100-1}):
\begin{equation}
\frac{1}{\bar{g}(t)^2} = \frac{1}{g^2} \left(\frac{F}{\mu^4}\right)^a.
\label{103}
\end{equation}
In terms of the parameter $t$ it follows that 
\begin{equation}
\bar{g}(t)^2= g^2 e^{-at}.
\label{104}
\end{equation}
Introducing these results in the effective Lagrangian we obtain
\begin{eqnarray}
{\cal L}_{eff} = \frac{1}{4g^2}F \left(\frac{F}{\mu^4}\right)^a.
\label{104+1}
\end{eqnarray}
For this case the function $G(\bar{g})$ results in
\begin{equation}
G(\bar{g}) = -\frac{1}{2}a.
\label{107}
\end{equation}
Thus since $a>0$ we conclude that $G(\bar{g})<0$, and again there will be no propagation
associated with the $e$-ray.
In the both cases, once the superluminal propagation is suppressed, 
the medium allows just one polarization mode to propagate. So it seems to behave like a polarizer.

\section{Observable for birefringence}
As shown in the previous section, birefringence effects
can occur in the Yang-Mills fields. Nevertheless, due
to confinement phenomenon, a direct measurement of these effects 
on gluons propagating in an external color field seems to be improbable.
However, deconfined quark-gluon matter, also known as quark-gluon plasma (QGP),
is expected to be produced in the Gold-Gold interaction at RHIC or 
Lead-Lead interaction at LHC.
This provides an opportunity to investigate the issue of gluon propagation 
in QGP systems. 

The different velocities in which the field disturbances can propagate
are associated with different polarization modes. It can be 
simply understood as a quantum measurement on the gluon, by which the gluon fall onto the
eigenstate of the polarization mode. In the special cases where the
$e$-ray gluon is forbidden by causality, the external field works
like a polarizer, and only the $o$-ray gluon is allowed there. The
polarization is assigned by the external field.
When one gluon propagates in a QGP, it is `measured' time and
time by the local field. Hence, the last polarization direction before
its hadronization into hadrons is completely out of control. This will
also destroy any  global polarization information \footnote{One
of the examples is that originated from the non-central
interaction, as suggested by the Shandong Group \cite{liangw}.
Corresponding to those cases, the phenomenon suggested here
appears as a `local polarization'. Hence this local polarization
destroys the information of the  global one and itself is not
measurable if we simply take the average in the experiment.
The STAR Collaboration of RHIC has recently measured the global 
polarization effect \cite{s1,s2}. From the above discussion, it 
seems very natural to adopt the negative results from STAR, still 
with confidence that both of the polarization effects can exist.}.
However, due to the complete polarization at any local area, the polarization correlation
will be very strong. A crucial point is how to identify two particles at the same
local area with  the same polarization. 
In order to explore this point let us consider, for simplicity, a
`gluon plasma'.
When the hard parton propagates in the medium, it
works as a source of small disturbances on the external field, and could lead to 
the emission of Cherenkov radiation. The $o$-gluon can exist within the Cherenkov cone,
but the $e$-gluon must be outside. If the gluon is not confined it can be measured
that the polarizations for two gluons inside the cone are parallel. 
More specifically $\hat{\epsilon}_1 \cdot \hat{\epsilon}_2=1$.
The experimental results depend not only on the polarization transfer from the gluon to a certain kind of
hadron in hadronization process, but also on the identification of the Cherenkov cone.
So, one suggestion is to measure the polarization correlation of two particles of the same kind
(2 vector mesons, 2 hyperons, etc.) with almost parallel momenta and within
the same jet cone (when the jet can be identified, as expected in LHC).
By studying the correlation dependent on the jet cone angle, one may have a way to measure
the Cherenkov angle.

One  experimental observable for the polarization correlation
can be suggested. This can be extracted from
the ideal case of two $\Lambda$ particles with the same polarization
$\vec{P}$, with $P=|\vec{P}|$ representing the  polarization rate.
The conventional way to measure $\vec{P}$ of a single $\Lambda$ is by measuring
the direction vector (denoted as $\hat{p}$) of the momenta of the daughter particles, 
e.g., proton or pion from the $\Lambda$ decay, at the rest frame of $\Lambda$.
Then the angular distribution,
\begin{equation}
\frac{dN}{d cos \theta} \sim 1+ \alpha \hat{p} \cdot \vec{P}=1+\alpha P cos \theta,
\label{distr}
\end{equation}
can give the information on the polarization. Here $\alpha$ is the hyperon decay parameter.
From this equation we see that if the direction of $\vec{P}$ is random, the average of all
$\Lambda's$ gives zero, then $P$ is not able to be measured. However, for the two $\Lambda's$
with the same polarization, in the  rest frame of {\it each} $\Lambda$, {\it respectively},
the direction vectors $\hat{p}_1$ and $\hat{p_2}$ can be measured.
Then we calculate the expectation value  $<\hat{p}_1 \cdot \hat{p}_2>$, which results in
\begin{equation}
P=3 \sqrt{<\hat{p}_1 \cdot \hat{p}_2>}/\alpha.
\label{expofp}
\end{equation}
Because $<\hat{p}_1 \cdot \hat{p}_2>$ is a SO(3) scalar, it does not depend on the direction
of polarization. Thus, it can be averaged for all the jets of all the (QGP) events in order to
get the scalar value of averaged polarization from Eq (\ref{expofp}).
In an experiment, the $\Lambda$ pair of the same jet \footnote{If the jet is not defined it
can be considered the near-parallel pairs of the same kind of hyperon.}
are expected to have the same polarization with larger probability, as discussed above.

\section{The effective geometry issue}
The effective theory approach has long been considered as a possible way to understand
confinement phenomena. One of the possible ways to investigate these phenomena consists in 
the construction of analogue models in which confinement would be related with an
event horizon formation, as it occurs in the black hole physics. Such a interpretation
was considered in \cite{castorina07,elbaz} (see also the references therein). 

The results discussed in this paper can be read in the context of the optical geometry.
In this context some comments are in order. Eqs.\ (\ref{+})--(\ref{-}) can be 
presented in the appealing form
\begin{equation}
g_{\pm}^{\mu\nu}k_\mu k_\nu=0,
\label{g1}
\end{equation}
where we defined the two symmetric contravariant tensors
\begin{eqnarray}
g^{\mu\nu}_+&=&\eta^{\mu\nu}-
\gamma F^{\lambda\mu}F^{\nu}{}_{\lambda},
\label{g+}
\\
g^{\mu\nu}_-&=&\eta^{\mu\nu}.
\label{g-}
\end{eqnarray}
The inverse symmetric tensor $g_{\mu\nu}$ is defined in such way that
$g^{\mu\alpha}g_{\alpha\nu}=\delta^\mu_\nu$.
Therefore, for each propagation vector $k_\nu$, the corresponding tensor $g^{\mu\nu}$
plays the role of an effective metric tensor. Indeed, $k_\nu$ is a light-like (or null) vector
with respect to the associated metric tensor. It can be shown \cite{delorenci02} that it also satisfies
a geodesic equation in terms of the Christoffel symbols $\Gamma^\alpha{}_{\!\mu\nu}$
associated with this metric. In this way we can refer to $k_\nu$ as a geodesic null 
vector with respect to the effective metric tensor $g^{\mu\nu}$. 
This geometric interpretation could be used in order to produce an analogue model for confinement
based on the possible formation of an event horizon. In this case, the requirement of causal 
propagation would assume a nontrivial role. Inside the hadron, where quark and gluons can interact and
propagate, the velocity of the field disturbances would be expected to be smaller than 1. 
On the other hand, if the vacuum outside the hadron is described 
by the effective Lagrangian, the larger than 1 velocity could be interpreted as an indication of
confinement, since no physical observable could propagate there.
%
%

\section{Conclusions}
In this paper the propagation of field disturbances was
investigated in the context of the effective Yang-Mills
Lagrangian. The  general dispersion relations for one-parameter
Lagrangians, Eqs. (\ref{+}) and (\ref{-}), were derived employing the method presented in
\cite{delorenci00,birula70}. It was shown that birefringence phenomena can 
occur. 

Let us remark some points. First, it should be stressed that 
the method depends on the effective Lagrangian as $L(F)$, so the conclusions are quite general.
Second, the assumption of causal propagation of the signals sets non-trivial  
constraint when exploring specific solutions. Finally, for the case of a 
deconfined quark gluon system, which naturally provides an
effective external field, the  birefringence phenomena with gluons and its local  polarization
effects are expected to be observed at RHIC and LHC by  measuring the strong local spin
correlations of various hadrons from QGP. The measurements of the spin correlations
suggested here, at the same time, can be useful in assigning the details
of the effective fields in QGP. These informations provide opportunities to develop
the effective Lagrangian framework and hence the better understanding of QCD.

\acknowledgments
This work is partially supported by the Brazilian research agencies
CNPq and FAPEMIG (V.A.D.L.); and National Natural Science Foundation of China
(NSFC) with grant No.10775090, and China Scholarship Council (CSC) (S.Y.L.).

\end{document}